\documentclass[aps,prl,twocolumn,groupedaddress]{revtex4}

\usepackage{graphics} 
\usepackage{graphicx}
\usepackage{dcolumn}
\usepackage{bm}
\everymath{\displaystyle}
\usepackage{amsmath}

\setcitestyle{super}

\begin{document}

\title{Resistance oscillations  of two-dimensional electrons in crossed electric and  tilted magnetic fields.} 

\author{William Mayer}
\author{Sergey Vitkalov}
\email[Corresponding author: ]{vitkalov@sci.ccny.cuny.edu}
\affiliation{Physics Department, City College of the City University of New York, New York 10031, USA}
\author{A. A. Bykov}
\affiliation{A.V.Rzhanov Institute of Semiconductor Physics, Novosibirsk 630090, Russia}
\affiliation{Novosibirsk State University, Novosibirsk 630090, Russia}

\date{\today}

\begin{abstract} 

Effect of $dc$ electric field on  transport of highly mobile 2D electrons  is studied in wide GaAs single quantum wells placed in titled magnetic fields. The study shows that  in perpendicular magnetic field resistance oscillates  due to electric field induced Landau-Zener transitions between quantum levels that corresponds to geometric resonances between cyclotron orbits and periodic modulation of electron density of states. Magnetic field tilt inverts these oscillations.  Surprisingly the strongest inverted oscillations are observed  at  a tilt corresponding to nearly absent modulation of the electron density of states in  regime of  magnetic breakdown of semiclassical electron orbits. This  phenomenon establishes an example of quantum resistance oscillations due to Landau quantization, which occur in  electron systems with a $constant$ density of states. 

\end{abstract}
 

\maketitle

The quantization of electron motion in magnetic fields generates a great variety of  fascinating transport phenomena observed in condensed materials.  Shubnikov-de Haas (SdH) resistance oscillations\cite{shoenberg1984} and   Quantum Hall Effect (QHE)\cite{qhe} are  famous examples related to the linear response of electrons. Finite electric fields produce remarkable nonlinear effects. At small electric fields Joule heating  strongly modifies   the 2D electron transport \cite{vitkalov2007,zhang2009,mamani2009,dmitriev2005} yielding exotic electronic states in which voltage (current) does not depend on current \cite{bykov2007zdr,zudov2008zdr,gusev2011zdr} (voltage\cite{bykov2013zdc}).   Application of a stronger electric field $E$ produces spectacular resistance oscillations.\cite{yang2002,bykov2005,zudov2007R,zudov2011,bykov2011,dietrich2012b} The oscillations are periodic with the electric field and obey  the following relation:
\begin{align}
\gamma eR_cE=j\hbar\omega_c,
\label{zener}
\end{align} 
where $e$ is electron charge, $R_c$ is the radius of cyclotron orbits of electrons at Fermi energy $E_F$, $j$ is a positive integer and factor $\gamma \approx 2$. These oscillations are related to impurity assisted Landau-Zener transitions between Landau levels titled by the electric field\cite{yang2002} and can be treated as geometrical resonances  between cyclotron orbits and spatially modulated density of states.\cite{vavilov2007,khodas2008}

2D electron systems with multiple populated subbands  exhibit additional quantum magnetoresistance oscillations.\cite{coleridge1990,leadley1992,bykov2008a,bykov2008b,bykov2008c,mamani2008,bykov2009,goran2009}  These magnetointersubband oscillations (MISO)   are due to an alignment between Landau levels from different subbands $i$ and $j$ with corresponding bottom energies $E_i$ and $E_j$.  The level alignment produces  resistance maximums  at the condition
\begin{align}
\Delta_{ij}=k\hbar \omega_c,
\label{miso_eq}
\end{align} 
 where $\Delta_{ij}=E_j-E_i$ and the index $k$ is a positive integer\cite{magarill1971,polyanovskii1988,raikh1994,raichev2008}. At a half integer $k$  Eq.(\ref{miso_eq}) corresponds to   resistance minimums occurring  at  nearly constant density of states (DOS) for  broad levels.\cite{raikh1994,raichev2008}

An application of in-plane magnetic field to the muti-subband systems creates significant modifications of electron spectra leading to  fascinating beating pattern of SdH oscillations and magnetic breakdown of semiclassical orbits\cite{bobinger1991,hu1992,harff1997,gusev2007,kumada2008,gusev2008,mueed2015}.   Recently it was shown that MISO are strongly modified by the in-plane magnetic field leading to a spectacular collapse of the beating nodes due to  magnetic breakdown\cite{mayer2016}.

In this paper we present investigations of the effect of the electric field on electron transport in  three-subband  electron systems placed in tilted magnetic fields.  The study reveals that the  in-plane magnetic field inverts the electric field induced resistance oscillations described by Eq.(\ref{zener}). The strongest inverted oscillations are observed   at the HF-MISO nodes  in the regime of  magnetic breakdown, in the $absence$ of the modulations of the density of states at the fundamental frequency $1/\hbar \omega_c$.  At these conditions  the dissipative resistance reaches a minimum value, which is smaller than the resistance at zero magnetic field. 

\begin{figure}[t]
\vskip -0.4cm
\includegraphics[width=\columnwidth]{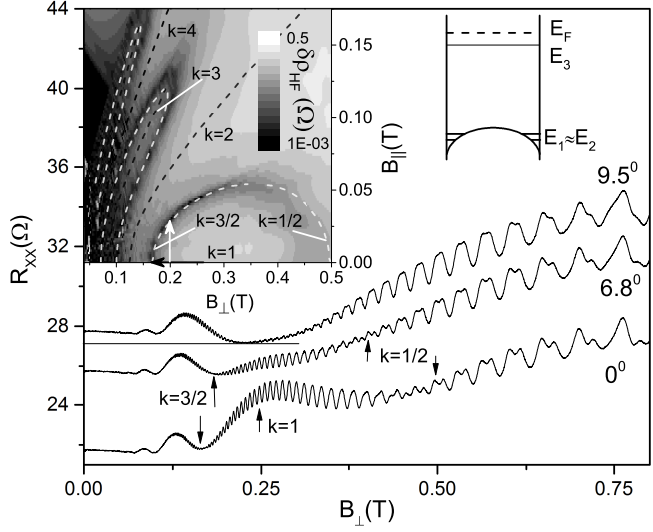}
\caption{Dependence of dissipative resistance on perpendicular magnetic field at different angles $\alpha$ as labeled. Curves are shifted for clarity. Right insert presents energy diagram of studied samples. Left insert presents  magnitude of HF-MISO in $B_\perp-B_\parallel$ plane. White (black) dashed lines present expected positions of HF-MISO nodes (LF-MISO maximums) obtained numerically\cite{mayer2016}. Sample A. }
\label{miso}
\end{figure}
 
Selectively doped GaAs single quantum well of width $d=$56 nm was grown by molecular beam epitaxy on a semi-insulating (001) GaAs substrate. The heterostructure has three populated subbands with energies $E_1\approx E_2<<E_3$ at the bottoms of the subbands. The energy diagram  are  schematically shown in the insert to Figure \ref{miso}.  Hall bars with width $W=50\mu$m ($y$-direction) and distance $L=250\mu$m ($x$-direction) between potential contacts demonstrating electron mobility $\mu \approx$1.6 $\times$10$^6$ cm$^2$/Vs and total density $n_T$=8.8$\times$ 10$^{15}$ m$^{-2}$ were studied at temperature 4.2 Kelvin. The magnetic field, $\vec  B$, was directed at different angles $\alpha$  relative  to normal to the samples and  perpendicular to the electric current.  Hall resistance  $R_H = B_\perp/(en_T)$ yields the angle $\alpha$, where $B_\perp=B\cdot cos(\alpha)$ is the perpendicular magnetic field.  Current $I_{ac}$=1$\mu$A at 133 Hz was applied through the current contacts and  the longitudinal and Hall $ac$ voltages ($V^{ac}_{xx}$ and $V^{ac}_H$) were measured in  response to a variable dc bias $I_{dc}$ applied through the same current leads.  The measurements were done in the linear regime in which the $ac$ voltages are proportional to  $I_{ac}$ yielding differential resistance $r_{xx}(I_{dc})=V^{ac}_{xx}/I_{ac}$.  Samples A and B with  slightly different gaps: $\Delta_{12}(A)$=0.43 meV and $\Delta_{12}(B)$=0.50 meV were studied.

Figure \ref{miso} presents a dependence of the  resistance $R_{xx}$ on the perpendicular magnetic field at different angles $\alpha$ as labeled. At $\alpha$=0$^o$ the resistance shows  low frequency (LF-MISO) and high frequency (HF-MISO) MISO.\cite{dietrich2015,wiedmann2010} LF-MISO correspond to the scattering between the two lowest symmetric (1) and antisymmetric (2 ) subbands and obey the relation $\Delta_{12}=k\hbar \omega_c$.\cite{mayer2016}  HF-MISO corresponds to scattering between either lowest and the third subband. Due to the  mismatch between  gaps: $\Delta_{13}-\Delta_{23}=\Delta_{12}$,  HF-MISO  show a beating pattern correlated with LF-MISO. In particular the nodes of HF-MISO beating are located at LF-MISO minimums. A parallel magnetic field, $B_\parallel$, moves nodes at $k$=1/2 and $k$=3/2 toward each other leading to  collapse at $\alpha$=9.5$^o$. Insert to Fig.\ref{miso} shows that odd $k$ LF-MISO maximums are bounded by the nodal lines.\cite{mayer2016} 
\begin{figure}[t]
\vskip -0.3cm
\includegraphics[width=\columnwidth]{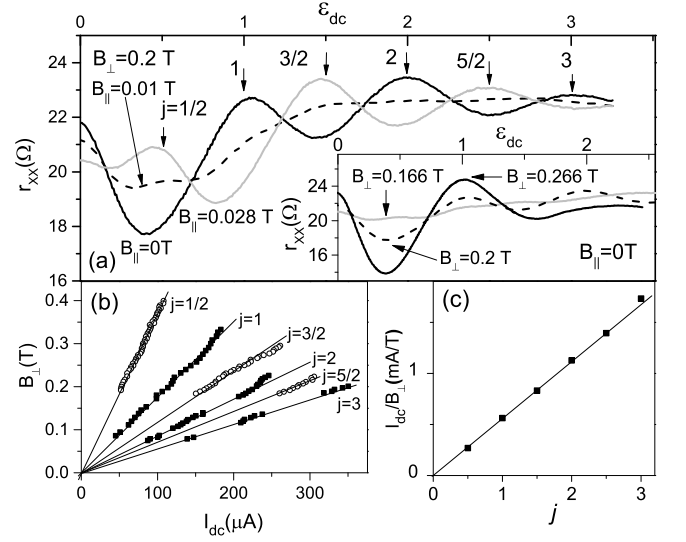}
\caption{(a)  Dependence of  differential resistance on normalized electric field, $\epsilon_{dc}=\gamma eR_c E/\hbar \omega_c$, where $\gamma$=1.9,  at different in-plane magnetic fields as labeled, obtained along the white arrow shown in Fig.\ref{miso}. Insert shows the resistance evolution along the black arrow shown in  the insert  to Fig.\ref{miso}; (b) Positions of resistance maximums  shown in (a)  at different magnetic fields $B_\perp$.  Lines present linear fit of the data; (c) Reciprocal slope of the linear fits shown in (b) vs index $j$ indicating agreement with Eq.(\ref{zener}). Sample A.  }
\label{zener1}
\end{figure} 

Figure \ref{zener1} presents   dependencies of the differential resistance $r_{xx}$ on the electric field $E$ at $B_\perp$=0.2 T and  different in-plane magnetic fields  as labeled  taken along the white arrow shown in the left insert to Fig.\ref{miso}.\cite{Ehall} At $B_\parallel$=0 T the black solid line  shows three maximums at $j$=1,2 and 3, which obey Eq.(\ref{zener}).  The gray  solid line presents the dependence taken at the end of the white arrow in the vicinity of the nodal line. This dependence  is inverted with respect to the black line and demonstrates maximums at $j$=1/2, 3/2 and 5/2. These maximums also obey Eq.(\ref{zener}) with the $same$ fundamental periodicity 1/$\hbar \omega_c$ but at the half integer values of the index $j$. The dashed line presents the dependence at an intermediate field, which does not display considerable oscillations.  
\begin{figure}[t]
\vskip -0.8cm
\includegraphics[width=\columnwidth]{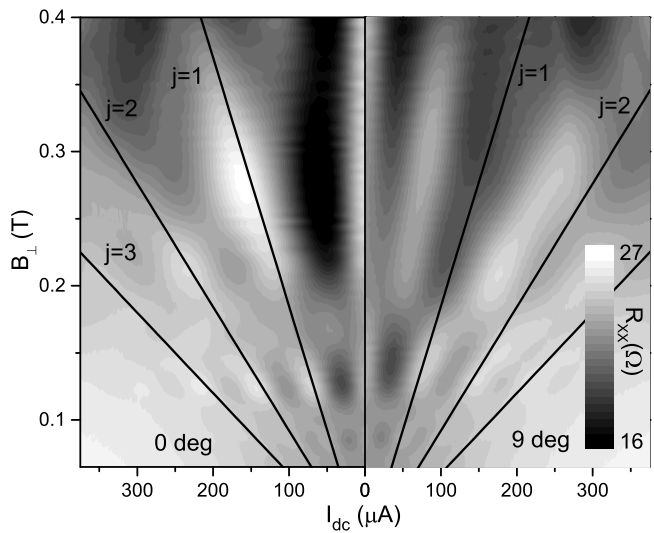}
\caption{ Dependence of differential resistance on dc bias and $B_\perp$ at two different angles as labeled. Solid lines present dependences obtained from Eq.(\ref{zener}) at $\gamma$=2 with no other fitting parameters. Sample A.}
\label{2D}
\end{figure} 
The insert to Fig.\ref{zener1} demonstrates the evolution of the electric field induced resistance oscillations taken along the black arrow shown in Fig.\ref{miso}. This evolution  is due to  variations of  the perpendicular magnetic field, $B_\perp$, at $B_\parallel$=0 T. These curves do not display an inversion. In contrast to the previous case at $k$=3/2 node the resistance oscillations  cease at the  fundamental frequency ($1/\hbar \omega_c$) and only  weak oscillations at second harmonics ($2/\hbar \omega_c$) are visible. This behavior is expected. Indeed in accordance with Eq.(\ref{miso_eq}) $k$=3/2 LF-MISO minimum and HF-MISO node correspond to the condition $\Delta_{12}=(3/2)\hbar \omega_c$. At this condition   symmetric and antisymmetric subband Landau levels  are shifted by 3/2$\hbar \omega_c$  with respect to each other and, therefore,  are equally spaced by $\hbar \omega_c/2$ near the Fermi energy. \cite{mayer2016} At $k$=3/2 the fundamental harmonic of the density of electron states (DOS) at frequency  $1/\hbar \omega_c$ is absent. Due to a small Dingle factor the amplitude of the second harmonic of  DOS  is exponentially small producing  very weak geometric resonances with cyclotron orbits at frequency $\sim 2/\hbar \omega_c$.\cite{dos}  The described behavior of the DOS is valid along all nodal lines\cite{mayer2016}  so the observed inversion of resistance oscillations is intriguing. 

The absence of the inversion at $B_\parallel$=0T suggests that the effect may have a relation to the magnetic breakdown of quasiclassical orbits.\cite{hu1992,harff1997,mayer2016,priestley1963,cohen1961,blount1962,slutskin1968,pippard1962,pippard1964} Figure \ref{2D} supports this proposal. The figure presents an overall behavior of the electric field induced resistance oscillations vs applied dc bias $I_{dc}$ and $B_\perp$ taken at two different angles.  At $\alpha$=0$^o$  magnetic breakdown is absent\cite{hu1992,mayer2016} and the  oscillations obey Eq.(\ref{zener}) with integer indexes $j$. Solid black lines present the theoretical dependence.\cite{yang2002,vavilov2007,khodas2008} The magnitude of the dc bias induce resistance oscillations is modulated by MISO.  At LF-MISO minimum $k$=3/2 ($B_\perp$=0.166 T) the oscillations are almost absent (see also insert to Fig.\ref{zener1}) and  are strongest in the vicinity LF-MISO maximums at $k$=1 and 2. While at angle $\alpha$=9.5$^o$ similar oscillations  are seen  in small $B_\perp$, the striking inversion of the oscillations is obvious at $B_\perp$$>$0.166 T.  Estimations indicate a 33\% probability of  magnetic breakdown at $B_\perp$=0.3 T and less than 3\%  at $B_\perp$$<$=0.166 T.\cite{hu1992,mayer2016} 
   
\begin{figure}[t]
\includegraphics[width=\columnwidth]{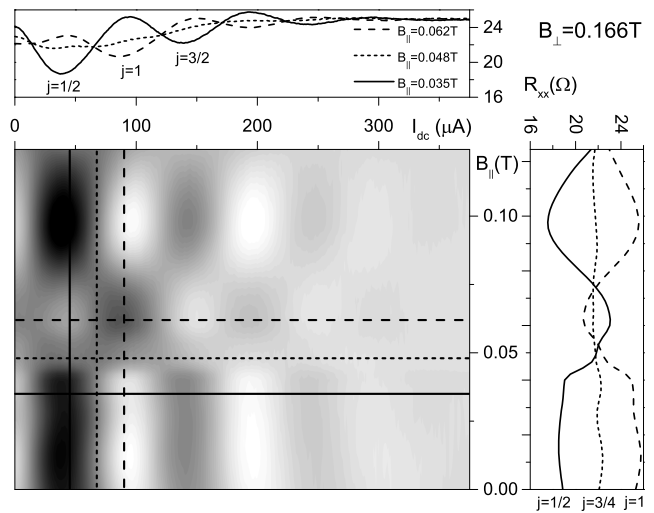}
\caption{Dependence of differential resistance on  dc bias and in-plane magnetic fields  at $B_\perp$=0.166 T.  Right panel shows re-inversion of dc bias induced oscillations with in-plane magnetic field. Sample B.  }
\label{rotation}
\end{figure}

Figure \ref{rotation} presents the evolution of the dc bias induced resistance oscillations for sample B taken in a vicinity of $k$=2 LF-MISO maximum at $B_\perp$=0.166 T and different $B_\parallel$. The obtained data demonstrate a re-inversion of the resistance oscillations suggesting  a periodicity of the inversion  with the in-plane magnetic field. Surprisingly oscillations of SdH amplitude in in-plane magnetic fields with a similar period  have been recently observed  (see Fig.8 in\cite{mayer2016}). These amplitude oscillations are related to periodic oscillations of the subband splitting $\Delta_{12}$ in strong magnetic fields.\cite{hu1992,moses1999,yakov2006,amro1,amro2} The right panel  indicates that at $j\approx$3/4  almost no resistance oscillations are induced by $B_\parallel$. The upper panel shows that this absence of  oscillations  holds at  $j\approx 1/4+p/2$, where $p$ is a positive integer.   

\begin{figure}[t]
\vskip -0 cm
\includegraphics[width=\columnwidth]{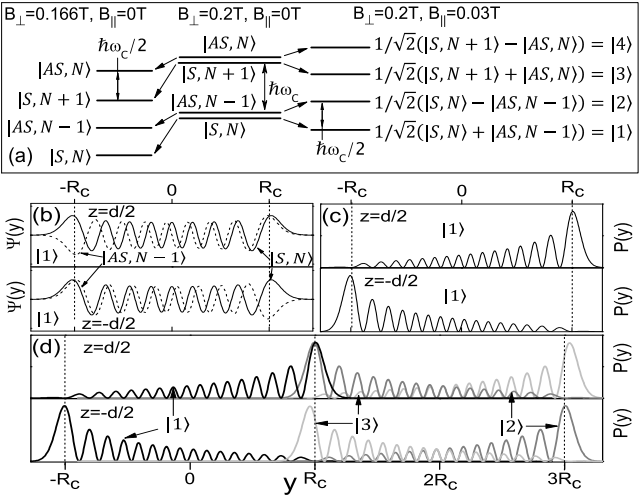}
\caption{(a) Evolution of energy spectra  due to  variation of cyclotron energy - left side and due to magnetic breakdown induced by  in-plane field- right side; (b) Eigenfunction $\vert 1\rangle$ presented  as linear combination of the basis set $\vert \xi,N\rangle$; (c)Spatial electron distribution in $\vert 1\rangle$ eigenstate in top  ($z$=$d/2$) and bottom ($z$=-$d/2$) 2D layers; (d)  Overlap between different eigenstates  during impurity backscattering.}
\label{theory}
\end{figure}

A theory of the observed inversion of  dc bias induced resistance oscillations is not available. Below a qualitative model  is proposed. Studied wide GaAs quantum wells  are considered as  two 2D parallel systems separated  by a distance $d$ in $z$-direction and  the coupling between the systems is treated in tight binding approximation using a  tunneling magnitude $t_0$.\cite{hu1992,mayer2016}  At  $B_\parallel$=0 T  electrons occupy symmetric (S) and antisymmetric(AS) subbands and move in $x-y$ plane along cyclotron orbits with radius $R_c$ at the Fermi energy. In $B_\perp$ the lateral electron motion is quantized and the eigenfunctions can be presented as $\vert \xi,N\rangle$, where $\xi$=S,AS and $N$=0,1,2... numerates Landau levels.\cite{mayer2016} An application of the in-plane magnetic field $B_\parallel \vert\vert E\vert\vert y$ mixes the symmetric and antisymmetric states. In the vicinity of the nodal line surrounding $k$=1 region  eigenfunctions are well approximated by a linear combination of one symmetric and one antisymmetric states (see Fig.10 in\cite{mayer2016}), which for  simplicity of the presentation we consider to be equally populated:  $\vert l\rangle$=($\vert S,N$+$1\rangle$$\pm$$\vert AS,N\rangle$)/$\sqrt 2$, where index $l$ numerates ascending energy levels. 
Figure \ref{theory}(a) presents an evolution of the electron spectrum  along the black and white arrows shown in Fig.\ref{miso}. The evolution corresponds to numerical computations of the spectrum in the vicinity of Fermi energy\cite{mayer2016}. 

Resistance oscillations are observed at high filling factors and, thus, the semiclassical treatment is appropriate.   It is accepted that the main contribution to  dc bias induced resistance oscillations comes from  electron backscattering by  impurities.\cite{yang2002,vavilov2007,khodas2008}   The backscattering occurs near  the turning points of the cyclotron orbits displaced by distance 2$R_c$ along the electric field $E$. The electron spends a considerable amount of time  at these points and the overlap between incident and scattered electron orbits is maximized.\cite{vavilov2007,khodas2008,efros2001,vitkalov2009} Below we analyze the spatial structure of eigenfunctions.    

Figure \ref{theory}(b) shows  the wave function $\vert 1\rangle$=($\vert S,N\rangle$+$\vert AS,N$-$1\rangle$)/$\sqrt 2$ for top  ($z$=$d/2$) and bottom ($z$=$-d/2$) 2D layers at $N$=16. Since  $N$ is even  the wave function $\vert S,N\rangle$ ($\vert AS,N$-$1\rangle$) is symmetric (antisymmetric) in both $y$ and $z$-directions. The eigenfunction $\vert 1\rangle$ is a sum  of these two functions that leads to the spatial electron distribution $P(y)=\vert\Psi(y)\vert^2$ shown in Fig.\ref{theory}(c): at the left (right) turning point of the oscillator state $\vert 1\rangle$ an electron is located mostly in the bottom (top) 2D layer at $-R_c$ ($R_c$). A similar configuration is obtained for  state $\vert 3\rangle$ while the electron distribution in state $\vert 2\rangle$ is the distribution in state $\vert 1\rangle$ rotated by 180$^o$ around the $y$=0 axes. 

The electric field $E$ tilts the spectrum  in $y$-direction (not shown) that allows horizontal transitions between the  levels due to elastic impurity scattering, which is considered as a local perturbation.\cite{yang2002,vavilov2007,khodas2008} The impurity backscattering near the turning points changes the direction of electron velocity by $\pi$, which is accomplished by an overlap between the incoming state near a  turning point and  the  outgoing state located near the $opposite$ turning point of the oscillator shifted by 2$R_c$ . Illustrating this statement Fig.\ref{theory}(d) indicates that the wave functions of the states $\vert 1\rangle$ and $\vert 2\rangle$ overlap at the opposite turning points, which  leads to  backscattering  while the backscattering between states $\vert 1\rangle$ and $\vert 3\rangle$ is significantly suppressed since these wave functions at the opposite turning points are located in $different$ 2D layers and, thus,  the overlap between two functions is exponentially small. Similar consideration  indicates the presence (absence) of backscattering  between  states $\vert l\rangle$ and $\vert m\rangle$ with  different  (the same)  parity of indexes:  $mod_2(m-l)=1$ ($mod_2(m-l)=0$). At nodal lines the energy difference between states with different index parity obeys the relation: $\delta E=E_m-E_l=\hbar\omega_c(j+1/2)$, that leads to the relation: $\gamma eR_c E=\hbar \omega_c(j+1/2)$ for the electric field induced resistance oscillations in  tilted magnetic field. 

At zero dc bias the backscattering occurs inside the same quantum level. Thus in tilted magnetic fields the impurity backscattering in the linear response is suppressed at the nodal lines since the parity of the incoming and outgoing states is the same. This conclusion is in agreement with the experiment. Indeed Fig.\ref{miso} shows that at the $k=3/2$ HF-MISO node located at $B_\perp$=0.2T and $B_\parallel$=0.033 T  the resistance reaches a value which is $less$ than the  value of the resistance both at $k$=3/2 at $B_\parallel$=0 T and even at zero magnetic field. The data indicates that  electron backscattering by impurities is effectively $controlled$ by  in-plane magnetic field. This result may have important implications for the field of topological insulators, where  electron backscattering is considered to be crucial.  

In conclusion the electric field induced resistance oscillations are studied in wide GaAs quantum wells placed in tilted quantizing magnetic field. The oscillations are related to impurity assisted Landau-Zener transitions between quantum levels and in perpendicular magnetic fields  obey relation: $2eR_cE=j\hbar \omega_c$, where $j$ is a positive integer. A tilt of the magnetic field inverts the oscillations. The strongest inversion occurs at the nodal line of the beating between magnetointersubband  resistance oscillations at which the density of electron states is nearly constant. These oscillations obey the relation $2eR_cE=j\hbar \omega_c$, where $j$ is a positive half integer. The effect is related to spatial redistribution  of  eigenfunctions of multi-subband  electron systems leading to $significant$ modification of the electron backscattering in tilted magnetic fields.

This work was supported by the National Science Foundation (Division of Material Research - 1104503),  the Russian Foundation for Basic Research (project no.14-02-01158) and  the Ministry of Education and Science of the Russian Federation.


\begin{thebibliography} {}
\bibitem{Ehall} Electric field $E$ equals  to Hall electric field, $E_H$ with accuracy  better than 1\% and the difference is ignored.
\end{thebibliography}
\end{document}